\documentclass[12pt]{article}
\usepackage{jheppub}
\usepackage[utf8]{inputenc}
\usepackage[T1]{fontenc} 
\usepackage{color, xcolor}
\usepackage{verbatim}
\usepackage{amsmath}
\usepackage{array}
\usepackage[normalem]{ulem}
\usepackage[hang,flushmargin]{footmisc}
\usepackage[verbose]{placeins}
\usepackage{appendix}

\pdfoutput=1

\newcommand{\be}{\begin{equation}}
\newcommand{\ee}{\end{equation}}
\newcommand{\bal}{\begin{equation}\begin{aligned}}
\newcommand{\eal}{\end{aligned}\end{equation}}
\newcommand{\bea}{\begin{eqnarray}}
\newcommand{\eea}{\end{eqnarray}}
\newcommand{\bi}{\begin{itemize}}
\newcommand{\ei}{\end{itemize}}

\def\ba#1\ea{\begin{align}#1\end{align}}
\def\bm#1\em{\begin{multline}#1\end{multline}}
\def\bmd#1\emd{\begin{multlined}#1\end{multlined}}
\def\a{\alpha}

\def\e{\epsilon}

\def\g{\gamma}

\def\k{\kappa}
\def\l{\lambda}

\def\m{\mu}
\def\n{\nu}
\def\p{\phi}
\def\P{\Phi}

\def\r{\rho}
\def\s{\sigma}

\newcommand{\la}{\label}

\newcommand{\re}{\ref}
\newcommand{\er}{\eqref}

\newcommand{\fr}{\frac}
\newcommand{\na}{\nabla}
\newcommand{\pa}{\partial}

\newcommand{\wtd}{\widetilde}

\newcommand{\eq}{\equiv}

\newcommand{\cd}{\cdots}

\newcommand{\qqu}{\qquad}

\renewcommand{\(}{\left(}
\renewcommand{\)}{\right)}
\renewcommand{\[}{\left[}
\renewcommand{\]}{\right]}
\newcommand{\<}{\langle}
\renewcommand{\>}{\rangle}

\newcommand{\cD}{{\mathcal D}}

\newcommand{\cL}{{\mathcal L}}

\newcommand{\cO}{{\mathcal O}}

\newcommand{\zb}{{\bar z}}

\newcommand{\ext}{\operatorname*{ext}}

\newcommand{\bulk}{\text{bulk}}

\newcommand{\gen}{\text{gen}}

\newcommand{\UV}{\text{UV}}
\newcommand{\IR}{\text{IR}}

\newcommand{\Tr}{\operatorname{Tr}}

\newcommand{\ignore}[1]{}

\title{Holographic entanglement from the UV to the IR}

\author[a]{Xi Dong,}
\affiliation[a]{Department of Physics, University of California,\\
Santa Barbara, CA 93106, USA}
\emailAdd{xidong@ucsb.edu}
\author[a,b,c]{Grant N. Remmen,}
\affiliation[b]{Kavli Institute for Theoretical Physics, University of California,\\
Santa Barbara, CA 93106, USA}
\affiliation[c]{Center for Cosmology and Particle Physics, Department of Physics, New York University,\\
New York, NY 10003, USA}
\emailAdd{grant.remmen@nyu.edu}
\author[a,d]{Diandian Wang,}
\affiliation[d]{Center for the Fundamental Laws of Nature, Harvard University,\\
Cambridge, MA 02138, USA}
\emailAdd{diandianwang@fas.harvard.edu}
\author[a,e]{Wayne W. Weng,}
\affiliation[e]{Department of Physics, Cornell University,\\
Ithaca, NY 14853, USA}
\emailAdd{www62@cornell.edu}
\author[a]{and Chih-Hung Wu}
\emailAdd{chih-hungwu@physics.ucsb.edu}

\begin{document}

\abstract{In AdS/CFT, observables on the boundary are invariant under renormalization group (RG) flow in the bulk. In this paper, we study holographic entanglement entropy under bulk RG flow and find that it is indeed invariant. We focus on tree-level RG flow, where massive fields in a UV theory are integrated out to give the IR theory. We explicitly show that in several simple examples, holographic entanglement entropy calculated in the UV theory agrees with that calculated in the IR theory. Moreover, we give an argument for this agreement to hold for general tree-level RG flow. Along the way, we generalize the replica method of calculating holographic entanglement entropy to bulk theories that include matter fields with nonzero spin.}

\maketitle

\section{Introduction}

Holographic entanglement entropy has played a significant role in understanding the emergence of spacetime in AdS/CFT, a duality between a bulk gravitational theory in anti-de Sitter space (AdS) and a boundary conformal field theory (CFT). At leading order in the large-$N$ expansion, the entanglement entropy of a subregion in the CFT is related to a geometric quantity in the bulk. For Einstein gravity, this quantity is simply the area of an extremal surface, given by the Ryu-Takayanagi (RT) formula \cite{Ryu:2006bv,Ryu:2006ef,Hubeny:2007xt}: 
\begin{equation} \label{eq:RT}
    S =\ext_\gamma \frac{2\pi}{\kappa^2}\int \mathrm{d}^{d}y \,\sqrt{h}, \qquad \k^2 = 8\pi G_N,
\end{equation}
where the extremization is over all candidate codimension-two surfaces $\gamma$ with induced metric $h$ and satisfying certain homology constraints. This formula was derived by Lewkowycz and Maldacena (LM) \cite{Lewkowycz:2013nqa} using the gravitational path integral.

For more general bulk gravity theories, such as higher-derivative gravity, the boundary entanglement entropy is instead given by \cite{Dong:2013qoa,Camps:2013zua,Dong:2017xht}
\begin{equation}\label{eq:SAgen}
    S = \ext_\gamma A_\gen[\gamma],
\end{equation}
where $A_\gen$ is an entropy functional evaluated on $\g$, generalizing the area in \er{eq:RT}. This formula can be derived via a generalization of the replica method of LM to arbitrary bulk theories.\footnote{Nevertheless, we will continue to refer to this generalization as the LM method.} For Einstein gravity, $A_\gen$ reduces to the area of the surface \eqref{eq:RT} divided by $4G_N$. For the case of $f(\text{Riemann})$ theories, the corresponding entropy functional was derived in Ref.~\cite{Dong:2013qoa}. For instance, when the bulk theory includes a Ricci-squared correction, ${\cal L} = (R + \lambda R_{\mu\nu}R^{\mu\nu})/2\kappa^2$, $A_\gen$ is given by the integral of the quantity $1+\lambda(R_a^{\;\;a} - K_a K^a/2)+\cO(\l^2)$ over the extremizing surface, where $K_a$ is the trace of the extrinsic curvature along the directions (or two-dimensional space) orthogonal to $\gamma$.\footnote{Throughout, we will use $a,b,\ldots$ for directions in the normal bundle to the codimension-two surface $\gamma$, while we use $i,j,\ldots$ for directions within the surface and Greek letters $\m,\n, \dots$ for full spacetime Lorentz indices.} In general, $A_\gen$ is a local functional determined from the action of the bulk theory. We will work to leading order in the gravitational constant $G_N$ and therefore ignore quantum corrections from bulk fields, which are generally subleading.

Since the holographic entanglement entropy formula relates a quantity on the boundary to a quantity in the bulk, this raises the question of what happens if there are different bulk descriptions of the same boundary theory. For the holographic duality to be consistent, the different bulk descriptions must give the same result for the boundary quantity, i.e.,
\begin{equation}
    S = \ext_\gamma A_{\gen,1}[\gamma; \Phi_1] = \ext_\gamma A_{\gen,2}[\gamma; \Phi_2].\label{eq:SEEequal}
\end{equation}
Here $\Phi_1$ and $\Phi_2$ represent schematically two distinct sets of dynamical fields (including the metric) that are present in the two bulk descriptions of the theory and evaluated on-shell in $A_\gen$, whereas $A_{\gen,1}$ and $A_{\gen,2}$ are the corresponding holographic entropy functionals.
The descriptions encoded by $\Phi_{1,2}$ could represent any reorganization of the bulk path integral, e.g., two different field redefinitions of the bulk action or a repackaging in terms of auxiliary fields.

\begin{figure}[t]
\centering
    \includegraphics[width=0.75\textwidth]{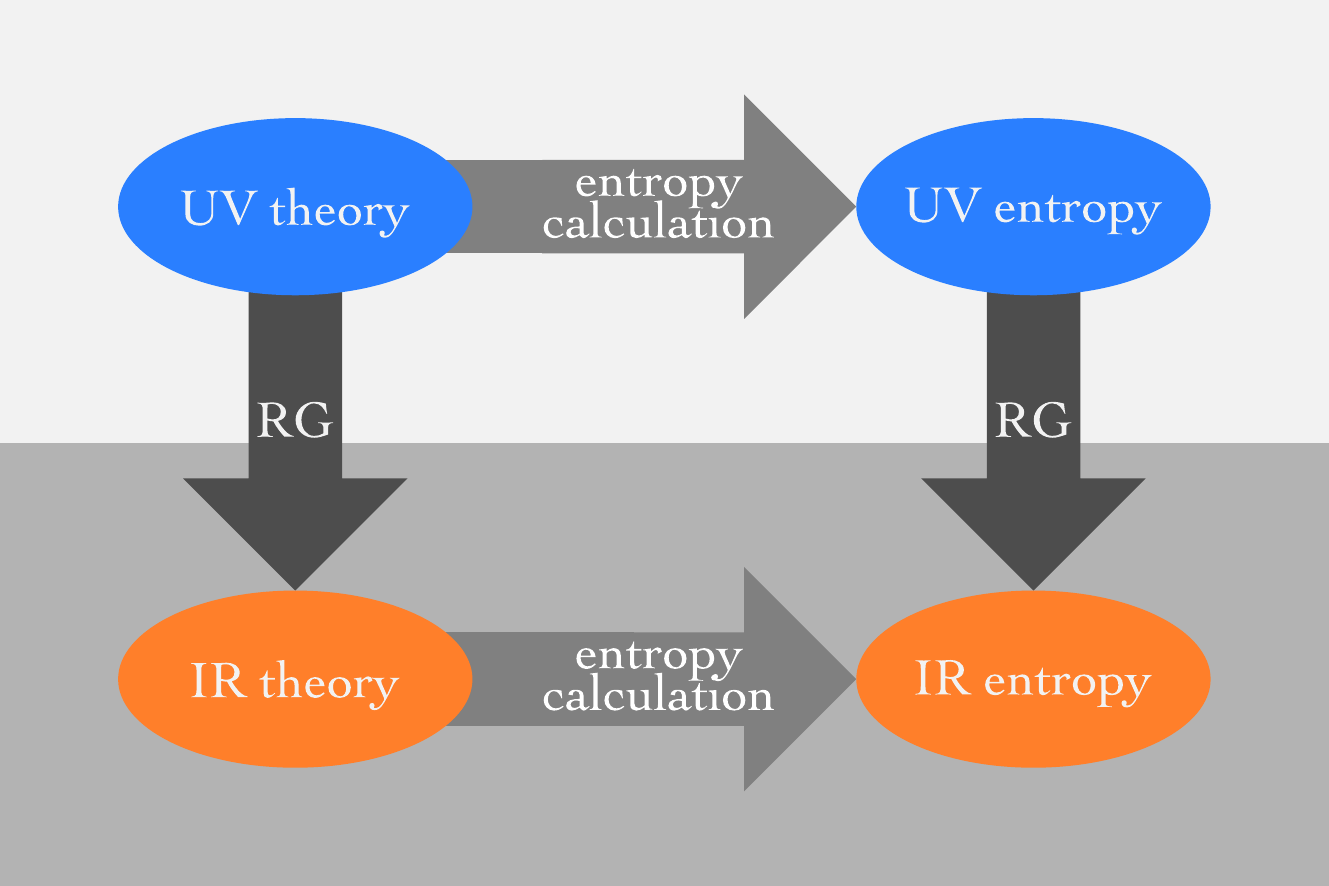}
    \caption{Commutative diagram for the holographic entanglement entropy calculation from the UV to the IR under bulk RG flow. The invariance of the entropy is demonstrated by first computing it in a tree-level UV extension with massive nonminimally coupled matter; upon integrating out the massive fields, the IR entropy agrees with that computed directly from the IR EFT.}
    \label{fig:commutative}
\end{figure}

One general situation where two bulk theories can have the same boundary description occurs in a bulk renormalization group (RG) flow, i.e., where the actions of $\Phi_{1,2}$ represent the same bulk theory evaluated at different energy scales. The boundary CFT is not sensitive to the scale at which the bulk effective field theory (EFT) is defined.\footnote{This is different from the boundary RG flow (which holographically corresponds to changing a radial cutoff in the bulk).} For this reason, Eq.~\eqref{eq:SEEequal} specializes to
\begin{equation}
    S = \ext_\gamma A_{\gen,\UV}[\gamma; g_{\rm UV}, \Phi] = \ext_\gamma A_{\gen,\IR}[\gamma; g_{\rm IR}],\label{eq:equalintro}
\end{equation}
where we focus on situations in which the IR theory is obtained by integrating out massive matter fields $\Phi$ in the UV theory, $\cL_\UV\[g_{\rm UV},\Phi\] \to \cL_\IR\[g_{\rm IR}\]$, and $A_{\rm gen,IR}$, $A_{\rm gen,UV}$ are the holographic entropy functionals in the IR and UV theory, respectively, with $g, \Phi$ evaluated on-shell.\footnote{Furthermore, we will find in explicit examples of gravitational EFTs and their tree-level UV extensions that a stronger version of Eq.~\eqref{eq:equalintro} holds, equating $A_{\gen,{\rm IR}}[g_\IR]$ with $\lim_{{\rm UV}\rightarrow {\rm IR}} A_{\gen,{\rm UV}}[g_\UV,\Phi]$ at the functional level---i.e., at the level of the entropy functional before extremization---where the ${\rm UV}\rightarrow {\rm IR}$ limit denotes imposition of the equations of motion for $\Phi$ at low energies.} Here, $g_\UV$ is a metric solution in the UV theory, and $g_\IR$ is the corresponding one in the IR. Requiring Eq.~\eqref{eq:equalintro} to hold is a nontrivial consistency check for the holographic entropy formulas.

Our main goal is to show that the UV entropy flowed to the IR matches the entropy computed in the IR theory, as summarized in Fig.~\ref{fig:commutative} as a commutative diagram. We do this by first demonstrating the matching in a few examples, namely certain UV extensions of four-derivative gravity, and then providing a general argument for all tree-level UV extensions of gravitational EFTs. 

Throughout, we work in the formalism of gravitational EFTs, where the leading corrections to the Einstein-Hilbert action are given by the four-derivative curvature-squared terms $R^2$, $R_{\mu\nu}R^{\mu\nu}$, and $R_{\mu\nu\rho\sigma}R^{\mu\nu\rho\sigma}$.
To study the RG flow in these theories, we introduce one-particle UV extensions, where the higher-derivative terms are generated at the tree level by integrating out massive fields nonminimally coupled to the curvature.\footnote{For example, in four dimensions one could couple a scalar $\phi$ of mass $m$ via the interaction term $M_{\rm Pl} \phi R$, generating $M_{\rm Pl}^2 R^2/m^2$ at low energies. In this way, the introduction of $\phi$ allows the four-derivative term to be replaced with one of lower mass dimension, while the scale at which the EFT breaks down is raised from $m$ to $M_{\rm Pl}$.} As a rule, we will refer to such theories as UV extensions rather than UV completions; what happens above their regime of validity will not affect our analysis. 

We emphasize that while we were initially inspired by the conjecture of Susskind and Uglum~\cite{Susskind:1994sm} which states that the generalized entropy---defined as the sum of $A_\gen$ and the bulk entropy $S_\bulk$---is invariant under RG flow, the statement that we want to make here is only about the behavior of $A_\gen$ itself. Indeed, as most of our discussion involves tree-level effects, we will confine our consideration to $A_{\rm gen}$ throughout. It would be interesting to investigate how our results extend to the full generalized entropy beyond the tree level.

The remainder of the paper is organized as follows. We start by specifying the tree-level extensions of four-derivative gravity in Sec.~\ref{sec:UV}, and show that the UV entropies match the entropies derived from the IR theory under the tree-level RG flow. The toolkit needed for deriving their entropy functionals is described in Sec.~\ref{sec:met}, with the details of the derivation presented in App.~\ref{sec:LM_RG}. In Sec.~\ref{sec:general}, we give a general argument for the entropy matching in tree-level UV extensions. We summarize our results and discuss certain future directions in Sec.~\ref{sec:disc}. In App.~\ref{sec:LM_redef}, we study holographic entanglement entropy under field redefinitions in two simple examples.

\section{Tree-level UV extensions and their entropy functionals} \label{sec:UV}

We now study holographic entanglement entropy under bulk RG flow in several simple examples.
 A physically motivated way to achieve such an RG flow is via a UV extension of the higher-derivative interactions where the equations of motion are second order in the UV.\footnote{Throughout, we use the phrase ``UV extensions'' rather than ``UV completions'' to denote integrating in massive fields that render the equations of motion second order and for which, upon integrating out these fields at the tree level, we obtain our desired EFTs. That is, we are replacing interaction terms of high mass dimension with ones of lower mass dimension.
In the case of massive scalars, this will raise the UV cutoff of the theory, while for massive higher-spin fields the situation is more subtle~\cite{Arkani-Hamed:2002bjr,Cheung:2019cwi}.
In all cases, we will be agnostic about the UV completion above the scale of the massive fields, as it will not play a role in our analysis.\label{foot:UV}}
We will therefore consider various tree-level UV extensions of the curvature-squared corrections $R^2$, $R_{\mu\nu}R^{\mu\nu}$, and $R_{\mu\nu\rho\sigma}R^{\mu\nu\rho\sigma}$.\footnote{We include $R^2$ and $R_{\mu\nu}R^{\mu\nu}$, even though IR EFTs with solely these corrections are equivalent to vacuum Einstein gravity under field redefinition, since we will find that working through the formalism demonstrating invariance of the holographic entropy under RG flow is already enlightening in these examples. We also discuss such field redefinitions in App.~\re{sec:LM_redef}.} 

Studying tree-level UV extensions is both well-motivated and tractable.
Tree-level completions in gravity arise naturally from a string-theoretic perspective; the nonzero effective curvature-squared terms that appear in (heterotic, type I, and bosonic) string theory~\cite{Gross:1986mw,Tseytlin:1995bi,Zwiebach:1985uq} are generated in the IR via a tower of massive states exchanged at the tree level between scattering gravitons.
Moreover, an $\hbar$-counting argument implies that graviton scattering in pure Einstein gravity must be UV-completed at the tree level in any weakly-coupled completion~\cite{CheungRemmen} (as opposed to a loop-level completion, as in the case of the Euler-Heisenberg Lagrangian), so considering higher-derivative gravitational terms that are themselves generated at the tree level may be well-motivated.
In an amplitude context, it has been shown that the K\"all\'en-Lehmann spectral representation describing the integrating out of massive degrees of freedom, even at the loop level, can be viewed as a sum over tree-level composite states~\cite{Arkani-Hamed:2020blm}.

As in Refs.~\cite{CheungRemmen,Cheung:2018cwt,Cheung:2019cwi}, we therefore integrate in new states at the tree level as a UV extension of the original EFT.\footnote{This is similar to the generation of a $(\partial\pi)^4$ term via $h(\partial\pi)^2$ in a linear sigma model of a Goldstone $\pi$ and Higgs field $h$~\cite{Adams:2006sv}.}  This allows us to access features of the RG evolution of the entropy calculation in toy models within the arena of quantum field theory and semiclassical gravity while sidestepping the full UV completion of gravity itself.

We now turn to the construction of our three example theories for generating the curvature-squared terms, using massive fields of spin zero, two, and four.

\subsection{Massive scalar extension of \texorpdfstring{$R^2$}{R2}}\label{sec:spinzeroUV}
To generate $R^2$, we consider a UV theory of a massive scalar nonminimally coupled to gravity, reminiscent of a dilaton in a model with broken supersymmetry.
The bulk action is
\begin{equation}
{\cal S} =\int{\rm d}^{d+2}x\,\sqrt{-g}\left[\frac{1}{2\kappa^{2}}R + \fr{\alpha}{\k}\phi R-\frac{1}{2}(\partial\phi)^{2} - \frac{1}{2}m^{2}\phi^{2}\right],\label{eq:action0}
\end{equation}
where $\alpha$ is a dimensionless constant. The $\phi$ equation of motion for this theory is 
\begin{equation}\label{eq:eoms}
\Box\phi-m^{2}\phi+\alpha\kappa^{-1}R = 0.
\end{equation}
At low energies, we find the solution
\be\la{phiIR}
\phi = \fr{\a R}{\kappa m^2}
\ee
up to higher order terms in $1/m^2$. We now integrate out $\phi$ at the tree level, plugging \er{phiIR} into \er{eq:action0} and finding the low energy EFT
\begin{equation}\label{eq:LEFT1}
\cL_{{\rm EFT}}=\frac{1}{2\kappa^{2}}R+\frac{\alpha^{2}}{2\kappa^{2}m^{2}}R^{2},
\end{equation}
dropping terms going like $(\alpha^2/m^2)\times \cO(1/m^2)$. 
The entropy functional for the UV theory~\er{eq:action0} is derived in App.~\ref{ssec:spin0hee} and is given by
\be
    A_\gen=\frac{2\pi}{\kappa^2} \int \mathrm{d}^dy\sqrt{h}\, \Big(1+2\alpha \kappa\phi\Big).\label{eq:Agenfinal0}
\ee
We note that this equation is exact to all orders in $\alpha$ and $1/m^2$. After RG flow to the IR---that is, taking the large-$m$ limit while keeping $\alpha/m$ fixed---the entropy functional $A_\gen$ becomes
\be \la{eq:IR_spin0}
    A_\gen=\frac{2\pi}{\kappa^2} \int \mathrm{d}^dy\sqrt{h}\, \left(1+\frac{2\alpha^2}{m^2} R\right).
\ee
This expression is again exact in $\alpha$, but has been truncated to leading order in the $1/m^2$ expansion, since we have flowed to the IR.
This result matches the entropy formula for the IR theory \eqref{eq:LEFT1}, computed using Eq.~\eqref{eq:S_Dong13} and following the prescription to be reviewed in Sec.~\ref{sec:met}. Thus, we have demonstrated entropy matching in this example.

\subsection{Massive spin-two extension of \texorpdfstring{$R_{\mu\nu}R^{\mu\nu}$}{Ricci2}}\label{sec:spintwoUV}

To generate the $R_{\mu\nu}R^{\mu\nu}$ term, we consider a UV theory that couples the Ricci tensor to a massive spin-two field $\phi_{\mu\nu}$. As noted in footnote~\ref{foot:UV}, this is not in general a full UV completion, and massive higher-spin states famously can lead to worse UV divergences or ghosts in the action unless we carefully choose the kinetic, mass, and interaction terms. 
Nonetheless, such considerations will not impact our purposes here, where we are interested in treating $\phi_{\mu\nu}$ effectively as a massive auxiliary field, providing us with a system that is equivalent to $R+R_{\mu\nu}R^{\mu\nu}$ gravity in the IR, while still possessing a second-order equation of motion that will make the entropy calculation qualitatively different.
As such, we will not require $\phi_{\mu\nu}$ to be a canonical (i.e., Fierz-Pauli) massive spin-two state~\cite{Hinterbichler:2011tt}.

Thus, we consider the UV action
\begin{equation}
{\cal S} =\int{\rm d}^{d+2}x\sqrt{-g}\left(\frac{1}{2\kappa^{2}}R+\fr{\alpha}{\k}\phi_{\mu\nu}R^{\mu\nu}-\frac{1}{2}\nabla_{\alpha}\phi_{\mu\nu}\nabla^{\alpha}\phi^{\mu\nu}-\frac{1}{2}m^{2}\phi_{\mu\nu}\phi^{\mu\nu}\right).
\label{eq:action2}
\end{equation}
The equation of motion for $\phi_{\mu\nu}$ is
\begin{equation}
\Box\phi_{\mu\nu}-m^{2}\phi_{\mu\nu}+\a\k^{-1}R_{\mu\nu}=0,\label{eq:eomphi}
\end{equation}
which at low energies $\na/m \ll 1$ leads to an effective Lagrangian with a Ricci-squared term,
\begin{equation}\la{eq:eftspin2}
\cL_{{\rm EFT}}=\frac{1}{2\kappa^{2}}R+\frac{\alpha^{2}}{2\kappa^{2}m^{2}}R_{\mu\nu}R^{\mu\nu},
\end{equation}
again dropping terms going like $(\alpha^2/m^2)\times \cO(1/m^2)$. The UV entropy formula is derived in App.~\ref{ssec:spin2hee} at leading order in a $1/m^2$ expansion, keeping $\a/m$ fixed. The result is
\be
A_\gen = \frac{2\pi}{\kappa^2} \int {\rm d}^d  y \sqrt{h} \left[ 1 + \a\k\p_a^a - \frac{\a^2}{2m^2} K_a K^a \right] + \cO\(\fr{\a^4}{m^4}\).\label{eq:Agenfinal2}
\ee
After RG flow to the IR (which, in this case, amounts to the replacement $\p_{\m\n} = \a R_{\m\n}/\k m^2$), $A_\gen$ becomes
\be \la{eq:IR_spin2}
A_\gen = \frac{2\pi}{\kappa^2} \int {\rm d}^d  y \sqrt{h} \left[ 1 + \frac{\alpha^{2}}{m^{2}} \left(R_a^a{} -\frac{1}{2}K_a{} K^a{}\right) \right] + \cO\(\fr{\a^4}{m^4}\),
\ee
which agrees with the IR entropy formula derived from Eq.~\er{eq:S_Dong13}.

\subsection{Massive spin-four extension of \texorpdfstring{$R_{\mu\nu\rho\sigma}R^{\mu\nu\rho\sigma}$}{Riem2}}\label{sec:spinfourUV}
To complete our consideration of gravitational EFTs at four-derivative order, we now turn to the Riemann-squared term.
Consideration of Riemann-squared is in particular necessary if we are to consider Gauss-Bonnet gravity and is nontrivial because, unlike $R^2$ and $R_{\mu\nu}R^{\mu\nu}$, it cannot be related to Einstein gravity via a field redefinition.
Using the tree-level generation of the Gauss-Bonnet term in the IR EFT of various string theories~\cite{Gross:1986mw,Tseytlin:1995bi,Zwiebach:1985uq} as motivation, we will follow Ref.~\cite{CheungRemmen} and consider a toy model in which the Riemann tensor is coupled to a single massive quantum field, providing a spin-four-like\footnote{We say ``spin-four-like'' since a canonical spin-four field is totally symmetric on its four Lorentz indices, which is incompatible with the symmetries of the Riemann tensor. Also, as in the case of the spin-two field, we will not be using a canonical kinetic or mass term, possibly leading to ghosts or tachyons, though our results will be insensitive to these subtleties.} generalization of the spin-zero and -two cases considered previously.

Suppose that in the UV we have a massive field $\phi_{\mu\nu\rho\sigma}$ possessing all
of the index symmetries of the Riemann tensor, $\phi_{\mu\nu\rho\sigma}=-\phi_{\nu\mu\rho\sigma}=-\phi_{\mu\nu\sigma\rho}=\phi_{\rho\sigma\mu\nu}$
as well as $\phi_{\mu[\nu\rho\sigma]}=0$,
with action
\begin{equation}
{\cal S}\,{=}\!\int \!{\rm d}^{d{+}2}x\sqrt{{-}g}\left(\frac{1}{2\kappa^{2}}R{+}\fr{\a}{\k}\phi_{\mu\nu\rho\sigma}R^{\mu\nu\rho\sigma}{-}\frac{1}{2}\nabla_{\alpha}\phi_{\mu\nu\rho\sigma}\nabla^{\alpha}\phi^{\mu\nu\rho\sigma}{-}\frac{1}{2}m^{2}\phi_{\mu\nu\rho\sigma}\phi^{\mu\nu\rho\sigma}\!\right)\!.\label{eq:action4}
\end{equation}
The equation of motion for $\phi_{\mu\nu\rho\sigma}$ is
\begin{equation}
\Box\phi_{\mu\nu\rho\sigma}-m^{2}\phi_{\mu\nu\rho\sigma}+\alpha\kappa^{-1}R_{\mu\nu\rho\sigma}=0,
\end{equation}
where, again in the regime $\na/m \ll 1$, we have a low-energy effective Lagrangian with a Riemann-squared term,
\begin{equation}
\cL_{{\rm EFT}}=\frac{1}{2\kappa^{2}}R+\frac{\alpha^{2}}{2\kappa^{2}m^{2}}R_{\mu\nu\rho\sigma}R^{\mu\nu\rho\sigma}, \label{eq:EFT4}
\end{equation}
up to terms that go like $(\alpha^2/m^2)\times \cO(1/m^2)$. The UV entropy formula is derived in App.~\ref{ssec:spin4hee}, again at leading order in a $1/m^2$ expansion, keeping $\a/m$ fixed. The result is
\be
A_\gen = \frac{2\pi}{\kappa^2} \int {\rm d}^d  y \sqrt{h} \left[ 1 +  2\alpha \kappa \phi^{ab}{}_{ab} -\frac{2\alpha^{2}}{m^{2}}K_{aij}{}K^{aij}{} \right]+\cO\left(\frac{\a^4}{m^4}\right).\label{eq:Agenfinal4}
\ee
After RG flow to the IR (using the replacement $\p_{\m\n\r\s} = \a R_{\m\n\r\s}/\k m^2$), $A_\gen$ becomes
\be \la{eq:IR_spin4}
A_\gen = \frac{2\pi}{\kappa^2} \int {\rm d}^d  y \sqrt{h} \left[ 1 + \frac{2\alpha^{2}}{m^{2}} \left(R^{ab}{}_{ab} -K_{aij} K^{aij}\right) \right] + \cO\(\fr{\a^4}{m^4}\),
\ee
which agrees with the IR entropy formula using Eq.~\eqref{eq:S_Dong13}.

\section{Holographic entropy toolkit}\label{sec:met}

In this section, we describe how to derive the entropy functionals presented in the previous section. We will begin with a review of the LM method~\cite{Lewkowycz:2013nqa}, and then explain the generalizations needed for it to work in our examples. We will apply this method to the example theories discussed in Sec.~\ref{sec:UV}, but since the details of the calculation are complicated, we relegate the full calculation to App.~\ref{sec:LM_RG}.

\subsection{Review of Lewkowycz-Maldacena}

Consider a spatial subregion $R$ in a quantum field theory and a density matrix $\r_R$ on $R$. The entanglement entropy of the state is defined as
\begin{equation}
    S = -\operatorname{Tr}\(\rho_R \log \rho_R \).
\end{equation}
This is often computed by taking the $n\to1$ limit of the R\'enyi entropy, 
\begin{equation}\label{eq:renyi}
    S_n = -\frac{1}{n-1} \log \Tr\(\rho^n_R\),
\end{equation}
a procedure referred to as the \textit{replica trick}. For integer $n \geq 2$, the R\'enyi entropy may be computed by
\be\label{eq:renyi_PI}
S_n = -\fr{1}{n-1} \log \left[\fr{Z_n}{(Z_1)^n}\right],
\ee
where $Z_n$ is the partition function of the quantum field theory on an $n$-fold cover. This partition function is often computed via a Euclidean path integral that prepares $n$ copies of the state $\r_R$.

We now apply this replica trick to holographic field theories, following the LM method. In that context, we use the holographic dictionary to translate Eq.~\eqref{eq:renyi_PI} into a bulk gravitational path integral,
\be
Z_n = \int \cD \P \, e^{-I\[\P\]},
\ee
integrating over all bulk fields (including the metric) with boundary conditions set by the $n$-fold cover. Here, $I$ is the Euclidean gravitational action in the bulk.
In the limit $G_N \to 0$, this gravitational path integral can be computed by a saddle-point approximation $Z_n = e^{-I\left[B_n\right]}$
evaluated on the dominant saddle $B_n$. If we further assume that $B_n$ preserves the boundary $\mathbb{Z}_n$ replica symmetry, then the action can be computed by a $\mathbb{Z}_n$ quotient, $I[\hat{B}_n] = I[B_n]/n$, thought of as the on-shell action of the orbifold spacetime $\hat{B}_n = B_n/\mathbb{Z}_n$. Applying this prescription to Eq.~\eqref{eq:renyi_PI}, we find the formula for the R\'enyi entropy,
\be
S_n =\fr{n}{n-1}\(I[\hat{B}_n]-I[\hat{B}_1]\).
\ee

\begin{figure}[t]
    \centering
    \includegraphics[width=0.95\textwidth]{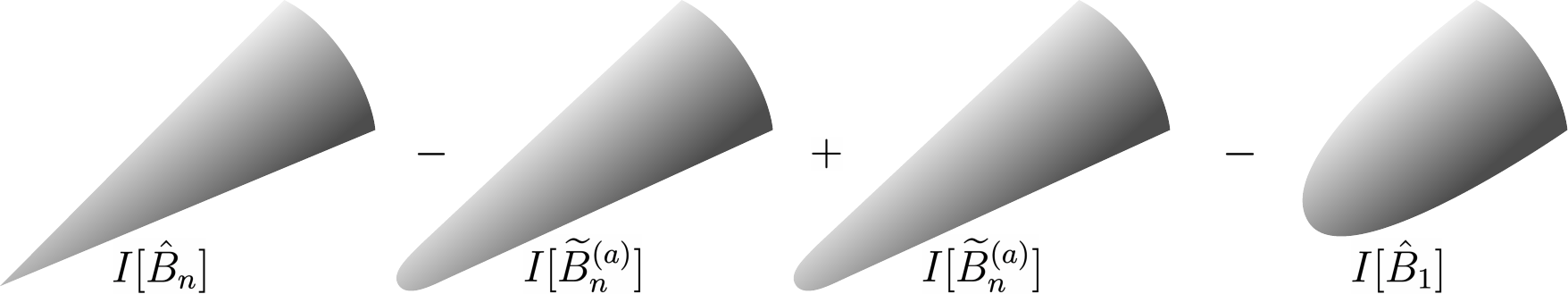}
    \caption{Pictorial representation of the regularized cone method, as detailed in Eq.~\eqref{eq:regu}. The leftmost picture represents the singular cone configuration $\hat{B}_n$, while the middle two pictures are its regularized version $\wtd{B}^{(a)}_n$. The rightmost picture is the solution $\hat{B}_1$ at $n = 1$.}
    \label{fig:LM_triple}
\end{figure}

The entanglement entropy is then obtained by taking the $n\to1$ limit, yielding
\begin{equation}
\begin{aligned}
    S = \left. \partial_n I[\hat{B}_n]\right|_{n=1}.
\end{aligned}
\end{equation}
To compute this quantity, we now use what is known as the regularized cone method~\cite{Lewkowycz:2013nqa}. We write 
\begin{equation} \la{eq:regu}
\begin{aligned}
    S = \lim_{n\to1}\frac{1}{n-1}\left(I[\hat{B}_n]-I[\wtd{B}_n^{(a)}]+I[\wtd{B}_n^{(a)}]-I[\hat B_1]\right),
\end{aligned}
\end{equation}
where we have subtracted and added the action of a new geometry, $\wtd{B}_n^{(a)}$, called the regularized cone, parameterized by a regulator $a$ and defined so that $\lim_{a\to0}\wtd{B}_n^{(a)}=\hat B_n$ and $\lim_{n\to 1}\wtd{B}_n^{(a)} = \hat{B}_1$ (independent of $a$); see Fig.~\ref{fig:LM_triple}.
The last two terms in Eq.~\er{eq:regu}  then cancel in the limit because $\hat B_1$ is on-shell. So the final expression is
\begin{equation}\label{eq:I0Ia}
S = \partial_n (I_0-I_a)\Big|_{n=1},
\end{equation}
where we have defined the shorthand $I_a\equiv I[\wtd{B}_n^{(a)}]$ and $I_0\equiv I[\hat B_n]$; note that $I_0$ can be obtained from $I_a$ by setting $a = 0$. Evaluating the variation in Eq.~\eqref{eq:I0Ia} ultimately leads to Eq.~\eqref{eq:SAgen}, where the entropy can be expressed as the extremum of an entropy functional $A_\gen$ that generally takes the form of an integral of a local quantity (e.g., a local geometric invariant)~\cite{Lewkowycz:2013nqa, Lewkowycz_2018, upcoming}.

The entropy functional has been derived using this method in the presence of higher-derivative terms. For example, for $\cL = f(\text{Riemann})$ gravity, it was derived in Ref.~\cite{Dong:2013qoa}:
\begin{equation}
\begin{aligned}
\label{eq:S_Dong13}
A_\gen
= &\; 2 \pi \int \mathrm{d}^{d} y \sqrt{h}\bigg\{-\frac{\partial \cL}{\partial R_{\mu \rho \nu \sigma}} \varepsilon_{\mu \rho} \varepsilon_{\nu \sigma}- \frac{\partial^{2} \cL}{\partial R_{\mu_{1} \rho_{1} \nu_{1} \sigma_{1}} \partial R_{\mu_{2} \rho_{2} \nu_{2} \sigma_{2}}} 2 K_{\lambda_{1} \rho_{1} \sigma_{1}} K_{\lambda_{2} \rho_{2} \sigma_{2}} \times \\
&\;\;\;{\times}\left[\left(n_{\mu_{1} \mu_{2}} n_{\nu_{1} \nu_{2}}{+}\varepsilon_{\mu_{1} \mu_{2}} \varepsilon_{\nu_{1} \nu_{2}}\right) n^{\lambda_{1} \lambda_{2}}{-}\left(n_{\mu_{1} \mu_{2}} \varepsilon_{\nu_{1} \nu_{2}}{+}\varepsilon_{\mu_{1} \mu_{2}} n_{\nu_{1} \nu_{2}}\right) \varepsilon^{\lambda_{1} \lambda_{2}}\right] {+} \cd\bigg\},
\end{aligned}
\end{equation}
where $\cL$ is the Lagrangian in Lorentzian signature, $K_{\lambda\rho\sigma}$ is the extrinsic curvature along the $x^\l$ direction, $\varepsilon_{\mu\nu}$ is the Levi-Civita tensor in the two normal directions (orthogonal to the surface), and $n_{\mu\nu}$ is the induced metric in the normal directions. The $\cdots$ represent terms involving higher orders in derivatives along the two normal directions. When evaluated in the examples that we studied in Sec.~\ref{sec:UV}, this expression precisely gives the IR entropies \eqref{eq:IR_spin0}, \eqref{eq:IR_spin2}, and \eqref{eq:IR_spin4}.

\subsection{Expansions in conical geometries}

We would now like to apply the LM method to our theories of interest and calculate the UV entropies.
Before doing so, several comments are in order. First, the derivation of the formula \eqref{eq:S_Dong13} was known to suffer to some extent from ambiguities called the splitting problem \cite{Dong:2013qoa, Miao:2014nxa, Dong:2017xht}. One aspect of this problem can be seen from the requirement that the desired expression for $A_\gen$ be a functional of quantities evaluated on a candidate codimension-two surface in the original spacetime $\hat{B}_1$, not the singular cone $\hat{B}_n$. 
To satisfy this requirement and resolve the ambiguities, we need to use certain equations of motion to express quantities in the singular cone in terms of $n=1$ quantities uniquely. 

A framework for solving the splitting problem concretely was provided in Ref.~\cite{Dong:2019piw}. It provided a prescription for a systematic \emph{conical expansion} of the metric, i.e., an expansion of the metric around the conical singularity when $n\ne1$. To begin, write the metric for the singular cone as in Ref.~\cite{Dong:2019piw}:
\begin{equation}
d s^2=d z d \bar{z}+T \frac{(\bar{z} d z-z d \bar{z})^2}{z \bar{z}}+h_{i j} d y^i d y^j+2 i U_j d y^j(\bar{z} d z-z d \bar{z}),\label{eq:sec3metric}
\end{equation}
where the coordinates $z,\zb$ satisfy the periodicity condition $z \sim z\,e^{2\pi i/n}$. For any $X\in\{T,U_i,h_{ij}\}$, expand $X$ in coupling constants $\lambda_1, \lambda_2, \dots, \lambda_k$ as 
\begin{equation}
    X = \sum_{r_1,\dots,r_k = 0}^\infty \l_1^{r_1} \cd \l_k^{r_k} X^{(\vec{r})}.\label{eq:Xexp}
\end{equation}
Additionally, the expansion coefficients $X^{(\vec r)}$ themselves have the following expansions in $z$ and $\zb$, 
\begin{equation}
\begin{aligned}
\label{eqs:triple}
    T^{(\vec{r})}&=\sum_{p, q=0,\, s= -r \atop p q>0 \text { or } s>0}^{\infty} T_{p q s}^{(\vec{r})} \,z^{np} \bar{z}^{nq}(z \bar{z})^{s},\\
        U_{i}^{(\vec{r})}&=\sum_{p, q=0,\, s=-r \atop p q>0 \text { or } s \geq 0}^{\infty} U_{i, p q s}^{(\vec{r})}\, z^{np} \bar{z}^{nq}(z \bar{z})^{s},\\
        h_{i j}^{(\vec{r})}&=\sum_{p, q=0,\, s=-r \atop p q>0 \text { or } s \geq 0}^{\infty} h_{i j, p q s}^{(\vec{r})}\, z^{np} \bar{z}^{nq}(z \bar{z})^{s},
\end{aligned}
\end{equation}
where the coefficients $X^{(\vec r)}_{pqs}$ are functions of $y^i$ only. We will refer to the expressions in Eq.~\eqref{eqs:triple} as \textit{conical expansions}.

Our goal here is to generalize the above prescription to theories involving additional matter fields, especially those nonminimally coupled to the metric. For a component $X$ of a general tensorial field (e.g., $X=\phi_{zi\zb j}$ or $X=R_{z\zb z\zb}$), we first define an expansion in terms of the coupling constants $\lambda_1, \l_2,\dots$ as in Eq.~\eqref{eq:Xexp}.
As in Eq.~\er{eqs:triple}, the coefficients $X^{(\vec r)}$ themselves have a conical expansion,
\be\la{Xtriple}
X^{(\vec r)} =\left(\frac{z}{\zb}\right)^{l/2}\sum_{p,q,s} X^{(\vec r)}_{pqs}\;  z^{np} \zb^{nq} (z\zb)^{s},
\ee
where the \textit{angular momentum} $l$ is defined to be the number of upper $z$ indices minus that of upper $\zb$ indices for the field component $X$ (each lower $z$ is considered as an upper $\zb$, and vice versa). The ranges of the indices $p,q,s$ depend on the specific theory and field component being considered. 

To find the entropy functional, we define a regularized metric given by
\be \label{eq:gregold}
ds^2 = e^{\e A} \[dz d\zb +\wtd T \fr{(\zb dz-z d\zb)^2}{z\zb}  + 2i \wtd U_j dy^j (\zb dz-z d\zb)\] + \wtd h_{ij} dy^i dy^j,
\ee
where $\wtd T$, $\wtd U_j$, and $\wtd h_{ij}$ are appropriately regularized versions of $T$, $U_j$, and $h_{ij}$, and $\e \equiv 1-1/n$.  When matter fields are present, we also regularize them in a similar way. Here $A$ is a suitable regulating function of $z,\zb$ and is parameterized by the same regulator $a$ that appeared in Eq.~\eqref{eq:regu}.
The entropy functional is found by evaluating Eq.~\er{eq:I0Ia} in this regularized field configuration. In App.~\ref{sec:LM_RG}, we show how to apply the above procedure to the examples in Sec.~\ref{sec:UV}.


\section{General argument for entropy matching}\label{sec:general}

In the examples we have considered in Sec.~\ref{sec:UV}, we found that the calculation of holographic entanglement entropy always commutes with RG flow. Specifically, whether we compute the entropy in the UV theory using the method of Sec.~\ref{sec:met} and then flow to the IR, or in the IR EFT directly using Eq.~\er{eq:S_Dong13}, we obtain the same result. This relationship is depicted in the commutative diagram shown in Fig.~\ref{fig:commutative}.

We now give a formal argument for this entropy matching in general theories.\footnote{It is worth noting that in the examples of Sec.~\ref{sec:UV}, matching works at the level of the entropy functional $A_\gen$ (before extremization). This is sufficient (although not necessary) for the entropy value (obtained after extremization) to match. In comparison, our general argument works at the level of the entropy value directly.}
We will show this result by proving that the R\'enyi entropies at integer R\'enyi index $n \geq 2$ satisfy this property. Our claim for the entanglement entropy then follows from the analytic continuation to $n = 1$.

To be concrete, consider a general IR EFT with a tree-level UV extension.
Recall from \er{eq:renyi_PI} that the R\'enyi entropy with integer $n \geq 2$ is computed from replicated partition functions.
Define $Z_{\rm UV}$ and $Z_{\rm IR}$ as the replicated partition functions of the UV extension and the IR EFT, respectively. 
By construction, the IR EFT is obtained by integrating out the UV degrees of freedom,
\be 
\int {\cal D}g \,e^{-I_{\rm IR}[g]} = \int {\cal D}g \,{\cal D}\phi \,e^{-I_{\rm UV}[g,\phi]},
\ee 
where we schematically write $\phi$ for any arbitrary UV fields, $g$ for the metric (and any other IR fields), and $I$ for the Euclidean action for each theory.
Thus, given well-defined boundary conditions for the metric (including the integer $n$ in the replica method), the partition functions are formally equal, $Z_{\rm IR} = Z_{\rm UV}$.
Hence, the R\'enyi entropies are equal. 
Therefore, their analytic continuation in $n$, and in particular the entanglement entropy, match between the UV and the IR, thus demonstrating our desired conclusion.

We can understand this result in more detail by investigating what happens near $n=1$, and more explicitly in terms of substituting in the equations of motion for the massive fields $\phi$.
The path integral for $\phi$ is dominated by its saddle point that solves the equations of motion (e.g., of the schematic form $(\Box -m^2)\phi \sim R$). Writing the solution collectively as $\bar\phi[g]$, we find at the tree level
\be
\int {\cal D}g \,e^{-I_{\rm IR}[g]} = \int {\cal D}g \,e^{-I_{\rm UV}[g,\bar\phi[g]]}.\label{eq:Ieq}
\ee
Now, both $I_{\rm IR}[g]$ and $I_{\rm UV}[g,\bar\phi[g]]$ in general contain an infinite tower of higher-curvature terms, which we can expand as a series in $\nabla/m$.
Retaining only the leading order in this series corresponds to the EFT requirement that all curvature scales are larger than the Compton wavelength of any UV states.
In this case, $\bar\phi[g]$ reduces to a local functional of $g$.
The path integral over $g$ is dominated by its saddle point, putting the metric on-shell, $g\rightarrow\bar g$. We therefore find
\be 
I_{\rm IR}[\bar g(n)] = I_{\rm UV}[\bar g(n),\bar\phi(n)],\label{eq:II}
\ee
where we explicitly write the argument $n$ to emphasize that the solutions are functions of the R\'enyi index $n$ defining the boundary conditions. Assuming that we can analytically continue in $n$ so that we can act with $\lim_{n\rightarrow 1}\partial_n$ on Eq.~\eqref{eq:II}, we obtain the matching of the von~Neumann entropy: on the left-hand side, it is the entropy computed directly in the IR EFT, while on the right-hand side, it is the entropy computed in the UV theory, which we RG-flow to the IR by coarse-graining the metric and field configuration on length scales much larger than $1/m$.

As we noted previously, in our example theories of Sec.~\ref{sec:UV} we made a stronger observation, namely that the entropy functionals in fact match (before extremization).
This can be understood by adding appropriate sources, as we will discuss in Sec.~\ref{sec:disc}.

\section{Discussion}\label{sec:disc}
In this paper, we have studied the matching of entanglement entropies between bulk gravitational effective theories and their UV extensions. 
In particular, through several example EFTs, we have demonstrated that the entropy computed directly in the IR theory matches that obtained by computing the entropy in the UV and flowing to the IR. We then gave a general argument based on the bulk gravitational path integral for why this matching should occur in any tree-level UV extension.

Along the way, we have generalized the LM method for deriving the holographic entropy formula to allow for certain matter fields. We expect this procedure to work for more general higher-derivative theories coupled to matter.

We now conclude with some open questions.

\paragraph{Entropy functional matching:} The argument in Sec.~\ref{sec:general} holds at the level of the entropy value $S$.
On the other hand, from the examples in Sec.~\ref{sec:UV}, we expect the matching to hold at the level of the entropy functional $A_\gen$, before imposing the extremality condition. Thus, it would be interesting to generalize the argument in Sec.~\ref{sec:general} to one that works at the level of the entropy functional.
One way of doing so is to allow the codimension-two surface to move from its extremal location to an arbitrary one. This can be accomplished by introducing a source term $T_{\mu\nu}\delta g^{\mu\nu}$ into the action, as in Ref.~\cite{Dong:2017xht}. Here, $T_{\mu\nu}$ is a background stress tensor giving a source for the metric fluctuation $\delta g_{\mu\nu}$, and it is chosen appropriately to allow us to open a conical defect along a general, not necessarily extremal, codimension-two surface.
We expect this to upgrade the argument in Sec.~\ref{sec:general} to one showing a stronger matching at the level of the entropy functional, although we leave the details to future work.\footnote{This functional-level matching between the UV and the IR is quite remarkable, as it need not hold in other contexts such as field redefinitions. In particular, as shown in App.~\ref{sec:LM_redef}, under field redefinitions the entropy values match, but not the entropy functionals in general.}

\paragraph{Finite-$\boldsymbol{m}$ entropy calculation:} For the spin-two and spin-four examples that we considered, in order to solve the equations of motion in the conical expansion, we found it convenient to perform a $1/m^2$ expansion in the UV. It would be interesting to see if it is possible to do a finite-$m$ calculation as in the scalar case.

\paragraph{Quantum corrections:}
In a more general context, we expect the full generalized entropy~\cite{Faulkner:2013ana, Engelhardt:2014gca}, rather than the value of $A_{\rm gen}$ alone, to match between the UV and IR as in Fig.~\ref{fig:commutative}.\footnote{See Ref.~\cite{Kabat:1995jq} for an example using an $O(N)$ invariant linear $\sigma$ model.} It would be interesting to extend our results to incorporate quantum corrections from bulk fields, especially to cases where the bulk matter entropy contribution is significant. Further, it would be illuminating to extend our argument to loop-level extensions of IR EFTs. We leave such investigations of quantum effects to future work.

\acknowledgments
We thank Don Marolf, Pratik Rath, and Zi-Yue Wang for interesting discussions, and especially Jiuci Xu for collaboration on related topics. This material is based upon work supported by the Air Force Office of Scientific Research under Award Number FA9550-19-1-0360. This material is also based upon work supported by the U.S. Department of Energy, Office of Science, Office of High Energy Physics, under Award Numbers DE-SC0023275 and DE-SC0011702. X.D. and W.W.W. were supported in part by funds from the University of California. G.N.R. is supported by the James Arthur Postdoctoral Fellowship at New York University, and was supported at the Kavli Institute for Theoretical Physics by the Simons Foundation
(Grant No.~216179) and the National Science Foundation (Grant No.~NSF PHY-1748958)
and at the University of California, Santa Barbara by the Fundamental Physics Fellowship.
D.W. is supported by NSF grant PHY2107939. C-H.W. was supported in part by the Ministry of Education, Taiwan.

\begin{appendix}

\section{Derivation of the entropy functionals}
\label{sec:LM_RG}
In this appendix, we compute the holographic entanglement entropy functional for each of the UV extensions given in Sec.~\ref{sec:UV} using the replica method described in Sec.~\ref{sec:met}.

Recall that in Sec.~\ref{sec:met}, the metric of the singular cone $\hat B_n$ with opening angle $2\pi/n$ can be written as in Eq.~\eqref{eq:sec3metric}, which we repeat here for convenience:
\be\label{eq:metsing}
ds^2 = dz d\zb +  T \fr{(\zb dz-z d\zb)^2}{z\zb} + 2i  U_j dy^j (\zb dz-z d\zb) + h_{ij} dy^i dy^j,
\ee
where the coordinates $z, \zb$ satisfy the periodicity condition $z \sim z \, e^{2\pi i/n}$. 
The regularized version of Eq.~\er{eq:metsing} is given in Eq.~\eqref{eq:gregold}, which we reproduce here as well:
\be\label{eq:metreg}
ds^2 = e^{\e A} \[dz d\zb +\wtd T \fr{(\zb dz-z d\zb)^2}{z\zb}  + 2i \wtd U_j dy^j (\zb dz-z d\zb)\]+ \wtd h_{ij} dy^i dy^j,
\ee
where as before $\e = 1-1/n$, and $A = A(a, z,\zb)$ is an appropriate regulating function that depends on the regulator $a$. A simple example that is sufficient for our calculation is
\be
A=\log \left(\frac{z^n\zb^n}{z^n\zb^n+a^{2n}}\right).
\ee
Recall that $\wtd T$, $\wtd U_j$, and $\wtd h_{ij}$, which appear in Eq.~\er{eq:metreg}, are appropriately regularized versions of $T$, $U_j$, and $h_{ij}$.
We denote the entire regularized metric defined by Eq.~\er{eq:metreg} as $\wtd g_{\m\n}$.
More generally, we write any regularized quantity with a tilde; for example, any matter fields $\p$ that are present in the theory are regularized to $\wtd \p$, and the regularized version of the Ricci scalar $R$ is written as $\wtd R$.
As in \er{Xtriple}, a field component $X$ (such as a component of $\phi$ or $R$) in the singular cone has a conical expansion with coefficients $X_{pqs}$, and its regularized version $\wtd X$ has a similar expansion with coefficients $\wtd X_{pqs}$.

Recall that the entropy $S$ can be calculated using Eq.~\er{eq:I0Ia} by evaluating the action difference between the singular and regularized cone to linear order in $\e$. Here we would like to derive the entropy functional $A_\gen$ whose extremum gives $S$. It can be shown that $A_\gen$ is given by an equation similar to Eq.~\er{eq:I0Ia}:
\be\la{eq:I0Ia2}
A_\gen = \pa_\e (I_0-I_a) \Big|_{\e=0},
\ee
where $I_0$ now denotes the action of a singular cone in which we only impose a suitable subset of equations of motion (EOMs)~\cite{upcoming}, and $I_a$ is an appropriately regularized version. Basically, not imposing all EOMs allows us to open a conical defect along a generic, not necessarily extremal, codimension-two surface, whereas imposing some of the EOMs allows us to solve quantities in the conical geometry in terms of quantities in the $n=1$ solution. Therefore, we will first evaluate Eq.~\er{eq:I0Ia2} as a function of conical expansion coefficients $X_{pqs}$ and then rewrite them in terms of standard Taylor expansion coefficients $X_{pq}$ in the $n=1$ solution:
\be
X \big|_{n=1} =\left(\frac{z}{\zb}\right)^{l/2}\sum_{p,q=0}^\infty X_{pq}\;  z^{p} \zb^{q},
\ee
by using the imposed EOMs and the continuity condition at $n=1$:
\be
X_{pq} = \sum_s X_{p-s,q-s,s} \Big|_{n=1}.
\ee
Here (and in the calculation below), we use commas to separate the indices of conical coefficients whenever necessary to avoid confusion. We will refer to $X_{pq}$ as the $n=1$ coefficients.

In principle, one can evaluate Eq.~\er{eq:I0Ia2} explicitly,
\begin{equation}\label{eq:full}
I_0-I_a=\int \(L_0 - L_a\),
\end{equation}
integrating over the entire bulk, where $L_0$ and $L_a$ are the Lagrangians (containing a factor of $\sqrt{g}$) evaluated on the singular and regularized cones, respectively.\footnote{We could denote $I_a$ and $L_a$ as $\wtd I$ and $\wtd L$, respectively, but for these two quantities we choose to use the current notation to make the $a$-dependence explicit.}
However, it can be shown~\cite{upcoming} that Eq.~\er{eq:full} reduces to
\begin{equation}\la{i0as}
    I_0-I_a=\int_{|z|<a} L_{0}-\int L_{a}^{\mathrm{I}}+\cO\big(\epsilon^{2}\big),
\end{equation}
and it is simpler to evaluate the right-hand side of this equation. Here $L_{a}^{\mathrm{I}}$ denotes the sum of a suitable subset of terms in $L_a$ that are called ``Type I'' terms \cite{upcoming}. Roughly speaking, these terms contain a factor of $\partial_z\partial_\zb A$ or its derivatives (after appropriate integration by parts), and they contribute to Eq.~\er{i0as} as regularized delta functions of $z, \zb$. The precise definition of Type I terms is provided in Ref.~\cite{upcoming}.
For simplicity, in this appendix we will use Eq.~\er{i0as} to calculate entropy functionals. However, one can always reproduce these results by using Eq.~\er{eq:full} directly, although the calculation would be more complicated.

\subsection{Massive scalar extension of \texorpdfstring{$R^2$}{R2}}\label{ssec:spin0hee}
Consider first the spin-zero theory \eqref{eq:action0} with Euclidean action
\begin{equation}
I=\int{\rm d}^{d+2}x\,\sqrt{g}\left[-\frac{1}{2\kappa^{2}}R-\fr{\alpha}{\k}\phi R+\frac{1}{2}(\partial\phi)^{2}+\frac{1}{2}m^{2}\phi^{2}\right].
\end{equation}
Since this theory has a scalar field, we need to prescribe an expansion for it in the singular cone similar to what we have done for the metric. As in the case of the metric \eqref{eqs:triple}, let us write the conical expansion for $\p$ as
\begin{equation}\la{phitriple}
\phi = \sum_{p,q,s=0}^\infty \phi_{pqs} \,z^{np}\zb^{nq} (z\zb)^s.    
\end{equation}
Due to the simplicity of this theory, we do not have to expand in $\a$ or $1/m^2$, so we will effectively work with finite $\a$ and $m$. In particular, the $s$ index in \er{phitriple} and the conical expansion of the metric is always nonnegative.

We now evaluate $I_0-I_a$ using Eq.~\er{i0as}. The only nonzero contribution comes from the $\wtd R_{z\zb\zb z}$ component in regularized Lagrangian $L_a$:
\be\la{eq:laterm}
    L_a \supset 2\(\sqrt{g}\)_{000}\(g^{z\zb}_{000}\)^2\partial_z\partial_\zb \(g_{z\zb,000}e^{\e A}\)\left(\frac{1}{2\kappa^2}+\frac{\alpha}{\kappa}\phi_{000}\right)(1+\cO(\e)).
\ee
Here, we have collected additional factors of $e^{\e A}$ into $(1+\cO(\e))$ because the entropy functional is not sensitive to anything beyond the linear order in $\e$ in the action. From now on, we will sometimes omit writing $(1+\cO(\e))$ in the regularized Lagrangian, with the caveat above understood. Using Eq.~\er{i0as}, we obtain the corresponding contribution to $I_0-I_a$ by integrating the $\pa_z \pa_\zb A$ term in Eq.~\er{eq:laterm}:
\begin{equation}
    I_0-I_a\supset 4\pi\e \int \mathrm{d}^dy \, (\sqrt{h})_{000}\left(\frac{1}{2\kappa^2}+\frac{\alpha}{\kappa}\phi_{000}\right)+\cO\big(\e^2\big).
\end{equation}

Using Eq.~\eqref{eq:I0Ia2} and the continuity condition at $n=1$
\begin{equation}
    \phi(z=0,\zb=0) \big|_{n=1} =\phi_{00} =\sum_{p=0}^{\infty}\phi_{p,p,-p} = \phi_{000},
\end{equation} 
the resulting entropy functional is given by
\be\label{eq:Agen_spin0}
    A_\gen=\frac{2\pi}{\kappa^2} \int \mathrm{d}^dy\sqrt{h}\, \big(1+2\alpha \kappa\phi\big),
\ee
which is precisely Eq.~\eqref{eq:Agenfinal0}. This formula is exact in the mass parameter $m$, i.e., we are not doing any large-mass expansion, in contrast to the spin-two and spin-four cases discussed in the next two subsections.

\subsection{Massive spin-two extension of \texorpdfstring{$R_{\m\n}R^{\m\n}$}{Ricci2}}\label{ssec:spin2hee}

In this subsection, we derive the entropy functional for the spin-two theory given in Eq.~\eqref{eq:action2}, with Euclidean action
\begin{equation}
I = \int{\rm d}^{d+2}x \sqrt{g} \( - \fr{1}{2\k^2} R -\fr{\a}{\k}\p_{\m\n} R^{\m\n} + \fr{1}{2}\na_\a\p_{\m\n}\na^\a \p^{\m\n} + \fr{1}{2}m^2 \p_{\m\n}\p^{\m\n}\).
\end{equation}
We work perturbatively in both $\a$ and $1/m^2$. In other words, we specialize Eqs.~\er{eq:Xexp} and \er{Xtriple} to
\be \la{eqs:exp}
X =\left(\frac{z}{\zb}\right)^{l/2}\sum_{r=0}^\infty\sum_{t=0}^\infty\sum_{p,q,s} X^{(r)\<t\>}_{pqs}\;\a^r m^{-2t} z^{np} \zb^{nq} (z\zb)^{s}. 
\ee
Here we have used a single $(r)$ index rather than $(\vec r)$ as it corresponds to one coupling constant $\a$ in this example. The angle brackets $\<\,\cdot\,\>$ label the order in a large-mass expansion as powers of $1/m^{2}$. 

Let us first analyze the equation of motion for $\phi_{\mu\nu}$ (which we impose):
\begin{equation}\label{eq:spin2EOM}
\Phi_{\m\n}\equiv\Box\phi_{\mu\nu}-m^{2}\phi_{\mu\nu}+\alpha\kappa^{-1}R_{\mu\nu}=0.
\end{equation}
Working at $\cO(\a^0)$, the equation at leading order in $1/m^2$ is
\begin{equation}
    \Phi^{(0)\<-1\>}_{\m\n}=-\phi^{(0)\<0\>}_{\mu\nu}=0.
\end{equation}
At the next order in $1/m^2$, we have
\be
    \Phi^{(0)\<0\>}_{\m\n}=\(\Box\phi_{\m\n}\)^{(0)\<0\>}-\phi^{(0)\<1\>}_{\m\n}=0.
\ee
Since $\phi_{\m\n}^{(0)\<0\>}=0$ from the previous order, this equation sets $\phi^{(0)\<1\>}_{\m\n}=0$. Repeating this procedure at each order in the large-mass expansion, we have 
\be\label{eq:spin2phi0}
\phi^{(0)}_{\m\n}=0
\ee
to all orders in $1/m^2$.

Moving to $\cO(\a)$, the matter equation of motion \eqref{eq:spin2EOM} at leading order in $1/m^2$ reads
\be \la{exp1-1}
    \Phi^{(1)\<-1\>}_{\m\n}=-\phi^{(1)\<0\>}_{\m\n}=0.
\ee
The next order in $1/m^2$ gives
\be
    \Phi^{(1)\<0\>}_{\m\n}=\(\Box\phi_{\m\n}\)^{(1)\<0\>}-\phi^{(1)\<1\>}_{\m\n}+\frac{1}{\k}R_{\m\n}^{(0)\<0\>}=0,
\ee
where the first term is zero according to Eq.~\er{exp1-1}, so
\be\label{eq:phi11}
    \phi^{(1)\<1\>}_{\m\n,pqs}=\frac{1}{\k}R_{\m\n,pqs}^{(0)\<0\>}.
\ee

We now evaluate $I_0-I_a$ using Eq.~\er{i0as}.
First, we work at $\cO(\a^0)$. Note that even though setting $\a=0$ in the Lagrangian gives Einstein gravity with matter, after using Eq.~\er{eq:spin2phi0} the Lagrangian is simply
\be
    L^{(0)}=-\frac{1}{2\k^2}(\sqrt{g})^{(0)}R^{(0)},
\ee
which is the same as in pure Einstein gravity without matter.
Similarly, the regularized version is
\be
    L_a^{(0)}=-\frac{1}{2\k^2}\(\sqrt{\wtd g}\)^{(0)}{\wtd R}^{(0)}.
\ee
Since the $s$ index of $\wtd R^{(0)}_{pqs}$ starts at $-1$, the only important contribution comes from the following term (in $\wtd R_{z\zb \zb z}$):
\be
    L_a^{(0)}\supset \frac{1}{\kappa^2}(\sqrt{g})^{(0)}_{000}\(g^{z\zb,(0)}_{000}\)^2\partial_z\partial_\zb \left(g^{(0)}_{z\zb,000}e^{\e A}\right).
\ee
This contribution integrates to
\begin{equation}\label{eq:s2L0einstein}
    (I_0-I_a)^{(0)}\supset \frac{2\pi\e}{\k^2} \int \mathrm{d}^dy \, \big(\sqrt{h}\big)^{(0)}_{000} + \cO(\e^2) = \frac{2\pi\e}{\k^2} \int \mathrm{d}^dy \, \big(\sqrt{h}\big)^{(0)}_{00} + \cO(\e^2),
\end{equation}
where we have rewritten the conical coefficient $\big(\sqrt{h}\big)^{(0)}_{000}$ in terms of the $n=1$ coefficient $\big(\sqrt{h}\big)^{(0)}_{00}$.
We identify the contribution of Eq.~\er{eq:s2L0einstein} as the zeroth-order area term in the series expansion in $\a$.

Next, at $\cO(\alpha)$, we find
\bal\label{eq:spin2L1}
\hspace{-1.5mm}L_a^{(1)}&=\(\sqrt{\wtd g}\)^{(1)}\wtd{\cL}^{(0)} \\
&\;\;+\!\(\sqrt{\wtd g}\)^{\!(0)}\!\!\bigg[{-}\,\frac{1}{2\k^2}\wtd R^{(1)}{-}\,\frac{1}{\k}\wtd R^{(0)}_{\m\n}\wtd \phi^{\m\n(0)}{+}\,(\wtd\nabla_\rho\wtd\phi_{\m\n})^{(0)}(\wtd \nabla^{\rho}\wtd \phi^{\m\n})^{(1)}{+}\,m^2\wtd \phi_{\m\n}^{(0)}\wtd \phi^{\m\n(1)}\bigg]\\
    &=\(\sqrt{\wtd g}\)^{(1)}\(-\frac{1}{2\k^2}\wtd R^{(0)}\)
    +\(\sqrt{\wtd g}\)^{(0)}\(-\frac{1}{2\k^2}\wtd R^{(1)}\),
\eal
where we have used Eq.~\eqref{eq:spin2phi0} to set $\phi_{\m\n}^{(0)}$ and its regularized version to zero. This is still just Einstein gravity, whose entropy functional is just the area, so we can write down its contribution directly:
\begin{equation}\label{eq:s2L1einstein}
    (I_0-I_a)^{(1)}=\frac{2\pi \e}{\k^2}\int \mathrm{d}^dy\,\[\sum_{p=0}^\infty (\sqrt{h})^{(1)}_{p,p,-p}\] + 
    \cO(\e^2) = \frac{2\pi \e}{\k^2}\int \mathrm{d}^dy\, (\sqrt{h})^{(1)}_{00} + 
    \cO(\e^2).
\end{equation}
Note that the above contribution comes solely from the first term on the last line of Eq.~\eqref{eq:spin2L1}, while the second term there turns out not to contribute, as can be checked explicitly.

At the next order $\cO(\alpha^2)$, we have
\bal\la{spin2L2}
  \hspace{-1mm}  L_a^{(2)}&= \Big(\sqrt{\wtd g}\Big)^{(2)}\[-\frac{1}{2\k^2}\wtd R^{(0)}\]
    +\Big(\sqrt{\wtd g}\Big)^{(1)}\[-\frac{1}{2\k^2}\wtd R^{(1)}\]\\
    &\;\;+\!\Big(\sqrt{\wtd g}\Big)^{\!(0)}\!\!\[{-}\frac{1}{2\k^2}\wtd R^{(2)}{-}\frac{1}{\k}\wtd R_{\m\n}^{(0)}\wtd \phi^{\m\n}{}^{(1)}
    {+}\frac{1}{2}(\wtd\nabla_\r\wtd\phi_{\m\n})^{(1)}(\wtd\nabla^\r\wtd\phi^{\m\n})^{(1)}{+}\frac{m^2}{2}\wtd\phi_{\m\n}^{(1)}\wtd\phi^{\m\n}{}^{(1)}\!\]\!.\hspace{-3mm}
\eal
The first three terms are from the Einstein-Hilbert action, so their contributions are known. In particular, the second and third terms turn out not to contribute, and the first term simply yields
\begin{equation}\label{eq:s2L2einstein}
    (I_0-I_a)^{(2)}\supset \frac{2\pi\e}{\k^2} \int \mathrm{d}^dy\,\[\sum_{p=0}^\infty (\sqrt{h})^{(2)}_{p,p,-p}\] + 
    \cO(\e^2) = \frac{2\pi \e}{\k^2}\int \mathrm{d}^dy\, (\sqrt{h})^{(2)}_{00} + 
    \cO(\e^2).
\end{equation}
This is similar to what we have seen at order $\cO(\a)$.

Let us now focus on the last three terms in \er{spin2L2} and work order by order in $1/m^2$. At the zeroth order in $1/m^2$, we find
\be
\hspace{-1mm}    L_a^{(2)\<0\>}
    {\supset}\! \(\!\!\sqrt{\wtd g}\)^{\!(0)\<0\>}\!\!\[\!{-}\frac{1}{\k}\wtd R_{\m\n}^{(0)\<0\>}\wtd \phi^{\m\n}{}^{(1)\<0\>}\!{+}
    \frac{1}{2}(\wtd \nabla_\r\wtd \phi_{\m\n})^{(1)\<0\>}\!(\wtd \nabla^\r\wtd\phi^{\m\n})^{(1)\<0\>}\!{+}\wtd \phi_{\m\n}^{(1)\<1\>}\wtd \phi^{\m\n}{}^{(1)\<0\>}\!\]\!.\hspace{-2mm}
\ee
This is set to zero by \er{exp1-1}. At first order in $1/m^2$, we have
\bal\label{eq:spin2-L21}
\hspace{-2mm}    L_a^{(2)\<1\>}\!\!&\supset
  \!  (\sqrt{\wtd g})^{(0)\<0\>}\!\!\[\!{-}\frac{1}{\k}\wtd R_{\m\n}^{(0)\<0\>}\wtd\phi^{\m\n}{}^{(1)\<1\>}\!{+}
    (\wtd\nabla_\r\wtd\phi_{\m\n})^{(1)\<0\>}\!(\wtd\nabla^\r\wtd\phi^{\m\n})^{(1)\<1\>}\!{+}\frac{1}{2}\wtd\phi_{\m\n}^{(1)\<1\>}\wtd\phi^{\m\n}{}^{(1)\<1\>}\!\]\\
    &=\!(\sqrt{\wtd g})^{(0)\<0\>}\!\!\[\!{-}\frac{1}{\k}\wtd R_{\m\n}^{(0)\<0\>}\wtd \phi^{\m\n}{}^{(1)\<1\>}+\frac{1}{2}\wtd \phi_{\m\n}^{(1)\<1\>}\wtd \phi^{\m\n}{}^{(1)\<1\>}\]. 
\eal

We now work out the contribution of Eq.~\er{eq:spin2-L21} to $I_0-I_a$ according to Eq.~\er{i0as}. First, focus on the first term in the last line of Eq.~\er{eq:spin2-L21}, which contributes in two ways. The first way is through the following Type I terms in $L_a$ coming from $(\mu,\nu)=(z,\zb)$ and $(\zb,z)$:
\bal
    L_a^{(2)\<1\>}\supset \(\sqrt{g}\)^{(0)\<0\>}_{000}\frac{1}{\k}g^{z\zb,(0)\<0\>}_{000}\partial_z\partial_\zb \left(g^{(0)\<0\>}_{z\zb,000}e^{\e A}\right)2\[\phi^{z\zb}{}^{(1)\<1\>}_{000}+\phi^{z\zb}{}^{(1)\<1\>}_{11-1}\],
\eal
which integrates to
\bal\label{eq:s2L21K0}
    (I_0-I_a)^{(2)\<1\>} \supset \int \mathrm{d}^dy\,\frac{2\pi\e}{\kappa} (\sqrt{h})^{(0)\<0\>}_{000}\, \[\phi^{z\zb}{}^{(1)\<1\>}_{000}+\phi^{z\zb}{}^{(1)\<1\>}_{11-1}\]+ \cO(\e^2).
\eal
Here $\phi^{z\zb}{}^{(1)\<1\>}_{11-1}$ could be written more clearly as $\phi^{z\zb}{}^{(1)\<1\>}_{1,1,-1}$, but we will use the former for simplicity.
The second way for the first term in Eq.~\er{eq:spin2-L21} to contribute is through the following non-Type I terms in $L_0$:
\bal
  L_0^{(2)\<1\>}\supset 
    \frac{1}{\k}\(\sqrt{g}\)^{(0)\<0\>}_{000}
    \bigg[&\,
    \frac{1}{2}h^{ij}{}^{(0)\<0\>}_{000}\pa_z\pa_z\( h_{ij}{}^{(0)\<0\>}_{100}z^{n}\)\phi^{zz}{}^{(1)\<1\>}_{01-1}
    \zb^{n-2}
    \\
    +&\,\frac{1}{2}h^{ij}{}^{(0)\<0\>}_{000}\pa_\zb\pa_\zb \(h_{ij}{}^{(0)\<0\>}_{010}\zb^{n}\)\phi^{\zb\zb}{}^{(1)\<1\>}_{10-1}
    z^{n-2}
    \bigg],
\eal
which is integrated within $|z|<a$ according to Eq.~\eqref{i0as} and gives
\bal
    (I_0-I_a)^{(2)\<1\>} \supset\int\mathrm{d}^dy\, (\sqrt{h})^{(0)\<0\>}_{000}\bigg[
    \frac{2\pi}{4\kappa} h^{ij}{}^{(0)\<0\>}_{000} h_{ij}{}^{(0)\<0\>}_{100}\phi^{zz}{}^{(1)\<1\>}_{01-1}+{\rm c.c.}  \bigg]+ \cO(\e^2).\label{eq:I0minusIa}
\eal
This expression is not manifestly $\cO(\e)$, but becomes so after we rewrite the spin-two field. Using the matter equations of motion \er{eq:phi11}, Eq.~\eqref{eq:I0minusIa} becomes
\bal
    (I_0-I_a)^{(2)\<1\>} &\supset\int\mathrm{d}^dy\, (\sqrt{h})^{(0)\<0\>}_{000}\bigg[
    \frac{2\pi}{4\kappa^2}h^{ij}{}^{(0)\<0\>}_{000} h_{ij}{}^{(0)\<0\>}_{100}R^{zz}{}^{(0)\<0\>}_{01-1}+{\rm c.c.}\bigg]+\cO(\e^2).\label{eq:I0minusIanew}
\eal
Recall the definition $K_{zij} \eq \fr12 \pa_z h_{ij}$. Using
\bal
    K_{z,00}{}^{(0)\<0\>}= \frac{1}{2}h^{ij}{}^{(0)\<0\>}_{00} h_{ij}{}^{(0)\<0\>}_{10} =\frac{1}{2}h^{ij}{}^{(0)\<0\>}_{000} h_{ij}{}^{(0)\<0\>}_{100}+\cO(\e),
\eal
and
\begin{equation}
\begin{aligned}
R^{zz}{}^{(0)\<0\>}_{01-1}
&=\,g^{z\zb}{}^{(0)\<0\>}_{000}g^{z\zb}{}^{(0)\<0\>}_{000}\[-\frac{1}{2}n(n-1)\]h_{ij}{}^{(0)\<0\>}_{010}h^{ij}{}^{(0)\<0\>}_{000}\\
&=-2n(n-1)h_{ij}{}^{(0)\<0\>}_{010}h^{ij}{}^{(0)\<0\>}_{000}\\
&=-4\e K_{\zb,00}^{(0)\<0\>}+\cO(\e^2),
\end{aligned}   
\end{equation}
Eq.~\eqref{eq:I0minusIanew} becomes
\be\label{eq:s2L21K2no1}
    (I_0-I_a)^{(2)\<1\>}
    \supset\int\mathrm{d}^dy\,(\sqrt{h})^{(0)\<0\>}_{000}\bigg[
    -\frac{4\pi\e}{\kappa^2}K_{z,00}{}^{(0)\<0\>}K_{\zb,00}^{(0)\<0\>}+{\rm c.c.}\bigg]+\cO(\e^2).
\ee
Now we move on to the second term in the last line of Eq.~\eqref{eq:spin2-L21}. It only contributes through the following non-Type I terms in $L_0$,
\bal
\hspace{-1.5mm}    \(L_0\)^{(2)\<1\>}\supset &\, \frac{1}{2}\(\sqrt{g}\)^{(0)\<0\>}_{000}\[\phi_{zz}{}^{(1)\<1\>}_{10-1}
    z^{n-2} 
    \phi^{zz}{}^{(1)\<1\>}_{01-1}
    \zb^{n-2}
    +\phi_{\zb\zb}{}^{(1)\<1\>}_{01-1}
    \zb^{n-2}
    \phi^{\zb\zb}{}^{(1)\<1\>}_{10-1}
    z^{n-2}
    \]\!,
\eal
which is again integrated according to Eq.~\eqref{i0as} and gives
\bal
    (I_0-I_a)^{(2)\<1\>}{\supset}\! \int \mathrm{d}^dy\,(\sqrt{h})^{(0)\<0\>}_{000}\[\frac{2\pi}{4\e} \phi_{zz}{}^{(1)\<1\>}_{10-1} \phi^{zz}{}^{(1)\<1\>}_{01-1}{+}{\rm c.c.}\]{+}\cO(\e^2).
\eal
Again using \er{eq:phi11}, this expression becomes
\bal\label{eq:s2L21K2no2}
    (I_0-I_a)^{(2)\<1\>}&{\supset}\!\int \mathrm{d}^dy\,(\sqrt{h})^{(0)\<0\>}_{000}\[\frac{2\pi}{4\k^2 \e} R_{zz}{}^{(0)\<0\>}_{10-1} R^{zz}{}^{(0)\<0\>}_{01-1}{+}{\rm c.c.}\!\]\!{+}\,\cO(\e^2)\\
    &{=}\!\int \mathrm{d}^dy\,(\sqrt{h})^{(0)\<0\>}_{000}\[\frac{2\pi\e}{\k^2}K_z{}^{(0)\<0\>}_{,00} K_\zb{}^{(0)\<0\>}_{,00}\times 2 \]{+} \,\cO(\e^2).
\eal

We note that Eqs.~\eqref{eq:s2L21K0}, \eqref{eq:s2L21K2no1}, and \eqref{eq:s2L21K2no2} are the only contributions of the last three terms in \er{spin2L2} to $I_0-I_a$ at $\cO(\a^2/m^2)$, as can be verified explicitly. Combining them gives 
\be 
\begin{aligned}
&\hspace{-1mm} (I_0\,{-}\,I_a)^{(2)\<1\>}\\
&\hspace{-1mm}\supset \, 2\pi\e \int \mathrm{d}^dy \,(\sqrt{h})^{(0)\<0\>}_{000}\bigg[
    \frac{1}{\kappa}\left(\phi^{z\zb}{}_{000}^{(1)\<1\>}
    {+} \phi^{z\zb}{}_{11-1}^{(1)\<1\>}\right)-\frac{2}{\kappa^2}K_{z,00}{}^{(0)\<0\>}K_{\zb,00}^{(0)\<0\>}\bigg]+\cO(\e^2).
\end{aligned}
\ee 
Adding this to the contributions at order $\a^0$ given in Eq.~\eqref{eq:s2L0einstein}, at order $\a$ given in Eq.~\eqref{eq:s2L1einstein}, and at order $\a^2$ given in Eq.~\eqref{eq:s2L2einstein}, we use Eq.~\eqref{eq:I0Ia2} to find the entropy functional
\be\label{eq:Agen_spin2}
A_\gen\,{=}\,\frac{2\pi}{\kappa^2} \int \! \mathrm{d}^dy\sqrt{h} \left[1\,{+}\,\frac{\alpha^2}{m^2}\left(\kappa\phi^{z\zb}{}^{(1)\<1\>} {-}{2}K_z{}^{(0)\<0\>}K_{\zb}{}^{(0)\<0\>}\right)\right]\!{+}\,\cO\(\fr{\a^4}{m^4}\)
\ee
up to higher order terms in $1/m^2$.
Here we have used the continuity condition at $n=1$
\be
    \phi_{\m\n}(z=0,\zb=0)\big|_{n=1}=\phi_{\m\n,00} = \sum_{p=0}^{\infty}\phi_{\m\n,p,p,-p}.
\ee
Writing Eq.~\er{eq:Agen_spin2} covariantly in $z$, $\zb$ and using $\phi^a_a{}^{(0)}=0$ from Eq.~\er{eq:spin2phi0}, we find
\be
A_\gen = \frac{2\pi}{\kappa^2} \int {\rm d}^d  y \sqrt{h} \left[ 1 + \alpha \kappa \phi_a^a{} -\frac{\alpha^{2}}{2m^{2}}K_a{}K^a{} \right]+\cO\(\fr{\a^4}{m^4}\),
\ee
which is precisely Eq.~\eqref{eq:Agenfinal2}.

\subsection{Massive spin-four extension of \texorpdfstring{$R_{\m\n\r\s}R^{\m\n\r\s}$}{Riem2}}\label{ssec:spin4hee}
Finally, we consider the spin-four theory \eqref{eq:action4} with Euclidean action
\begin{equation}
 I \,{=}\! \int{\rm d}^{d+2}x \sqrt{g} \(\!{-} \fr{1}{2\k^2} R {-}\fr{\a}{\k}\p_{\m\n\r\s} R^{\m\n\r\s} {+} \fr{1}{2}\na_\a\p_{\m\n\r\s}\na^\a \p^{\m\n\r\s} {+} \fr{1}{2}m^2 \p_{\m\n\r\s}\p^{\m\n\r\s}\!\)\!,
\end{equation}
and we will again work perturbatively in both $\a$ and $1/m^2$.

Repeating the order-by-order analysis that we performed for the spin-two field's equation of motion, we find analogously that, to all orders in the $1/m^2$ expansion,
\be\label{eq:spin4phi0}
\phi^{(0)}_{\m\n\r\s}=0.
\ee
We also find
\be 
\begin{aligned}
    \phi^{(1)\<0\>}_{\m\n\r\s,pqs} &=0,\\
    \label{eq:phi11s4}
    \phi^{(1)\<1\>}_{\m\n\r\s,pqs} &=\frac{1}{\k}R_{\m\n\r\s,pqs}^{(0)\<0\>}.
    \end{aligned}
\ee

We now analyze the action difference $I_0-I_a$ to linear order in $\e$. First, note that at $\cO(\a^0)$ the action is exactly the same as that of the spin-two theory with $\a$ set to zero, so we obtain Eq.~\eqref{eq:s2L0einstein} again without having to repeat the calculation. In other words, the leading-order result is just the area term of Einstein gravity. 

At $\cO(\alpha)$, using Eq.~\eqref{eq:spin4phi0}, we can write down the analogue of Eq.~\eqref{eq:spin2L1}:
\be\label{eq:spin4L1}
L_a^{(1)}=
    \(\sqrt{\wtd g}\)^{(1)}\(-\frac{1}{2\k^2}\wtd R^{(0)}\)
    +\(\sqrt{\wtd g}\)^{(0)}\(-\frac{1}{2\k^2}\wtd R^{(1)}\).
\ee
The second term turns out not to contribute (as in the spin-two case), and the first term gives Eq.~\er{eq:s2L1einstein} again.

At $\cO(\a^2)$, we have
\bal
    L_a^{(2)}&=\;\; \Big(\sqrt{\wtd g}\Big)^{(2)}\[-\frac{1}{2\k^2}\wtd R^{(0)}\]
    +\Big(\sqrt{\wtd g}\Big)^{(1)}\[-\frac{1}{2\k^2}\wtd R^{(1)}\]\\
    &\;\;\;+\Big(\sqrt{\wtd g}\Big)^{(0)}\bigg[-\frac{1}{2\k^2}\wtd R^{(2)}-\frac{1}{\k}\wtd R_{\m\n\r\s}^{(0)}\wtd \phi^{\m\n\r\s}{}^{(1)}\\
    &\qquad\qquad \;\;\;\;\;\;\,\,+\frac{1}{2}(\wtd\nabla_\a\wtd\phi_{\m\n\r\s})^{(1)}(\wtd\nabla^\a\wtd\phi^{\m\n\r\s})^{(1)}+\frac{m^2}{2}\wtd\phi_{\m\n\r\s}^{(1)}\wtd\phi^{\m\n\r\s}{}^{(1)}\bigg].
\eal
The first three terms give Eq.~\er{eq:s2L2einstein} again.
Now we focus on the remaining three terms and work order by order in $1/m^2$. At $\cO(m^0)$, they vanish after using the matter equations of motion.
At $\cO(1/ m^2)$, the remaining three terms give
\be\label{eq:spin4-L21}
    L_a^{(2)\<1\>}\supset (\sqrt{\wtd g})^{(0)\<0\>}\[-\frac{1}{\k}\wtd R_{\m\n\r\s}^{(0)\<0\>}\wtd \phi^{\m\n\r\s}{}^{(1)\<1\>}+\frac{1}{2}\wtd \phi_{\m\n\r\s}^{(1)\<1\>}\wtd \phi^{\m\n\r\s}{}^{(1)\<1\>}\].
\ee
The first term contributes through Type I terms
\bal
    (L_a)^{(2)\<1\>}\supset \(\sqrt{g}\)^{(0)\<0\>}_{000}\[\frac{1}{\k}\partial_z\partial_\zb \left(g^{(0)\<0\>}_{z\zb,000}e^{\e A}\right)4
    \(\phi^{z\zb\zb z}{}^{(1)\<1\>}_{000}+\phi^{z\zb\zb z}{}^{(1)\<1\>}_{11-1}\)
    \],
\eal
where the factor of 4 comes from symmetries of $\phi^{\m\n\r\s}$, and the above equation integrates to
\bal\label{eq:s4L21K0}
    (I_0-I_a)^{(2)\<1\>} \supset \frac{2\pi\e}{\kappa} \int \mathrm{d}^dy\,(\sqrt{h})^{(0)}_{000} \,\(\phi^{z\zb\zb z}{}^{(1)\<1\>}_{000}+\phi^{z\zb\zb z}{}^{(1)\<1\>}_{11-1}\) + \cO(\e^2).
\eal
The first term in Eq.~\er{eq:spin4-L21} also contributes through non-Type I terms
\bal
    \(L_0\)^{(2)\<1\>} \supset \,\frac{1}{\k}\(\sqrt{g}\)^{(0)\<0\>}_{000}\bigg[&\,
    \frac{1}{2}\pa_z\pa_z\( h_{ij}{}^{(0)\<0\>}_{100}z^{n}\)4\phi^{zizj}{}^{(1)\<1\>}_{01-1}
    \zb^{n-2}
    \\
    +&\,\frac{1}{2}\pa_\zb\pa_\zb \(h_{ij}{}^{(0)\<0\>}_{010}\zb^{n}\)4\phi^{\zb i\zb j}{}^{(1)\<1\>}_{10-1}
    z^{n-2}
    \bigg],
\eal
which is integrated according to Eq.~\er{i0as} and gives
\bal
    (I_0-I_a)^{(2)\<1\>} \supset\int\mathrm{d}^dy\, (\sqrt{h})^{(0)\<0\>}_{000} \bigg[
    \frac{2\pi}{\kappa} 
    h_{ij}{}^{(0)\<0\>}_{100}\phi^{zizj}{}^{(1)\<1\>}_{01-1}+{\rm c.c.}  \bigg]+ \cO(\e^2).\label{eq:I0minusIa_s4}
\eal
As before, we proceed by rewriting the spin-four field. Using the matter equations of motion \er{eq:phi11s4}, Eq.~\eqref{eq:I0minusIa_s4} becomes
\bal
    (I_0-I_a)^{(2)\<1\>} &\supset\int\mathrm{d}^dy\, (\sqrt{h})^{(0)\<0\>}_{000} \bigg[
    \frac{2\pi}{\kappa^2}h_{ij}{}^{(0)\<0\>}_{100}R^{zizj}{}^{(0)\<0\>}_{01-1}+{\rm c.c.}\bigg]+\cO(\e^2).\label{eq:I0minusIanew_s4}
\eal
Using
\bal
    K_{zij,00}^{(0)\<0\>}= \frac{1}{2}h_{ij,10}^{(0)\<0\>} =\frac{1}{2}h_{ij,100}^{(0)\<0\>} + \cO(\e)
\eal
and
\begin{equation}
\begin{aligned}
R^{zizj}{}^{(0)\<0\>}_{01-1}
&=\,g^{z\zb}{}^{(0)\<0\>}_{000}g^{z\zb}{}^{(0)\<0\>}_{000}\left(-\frac{1}{2}n(n-1)\right)h_{kl}{}^{(0)\<0\>}_{010}h^{ik}{}^{(0)\<0\>}_{000}h^{jl}{}^{(0)\<0\>}_{000}\\
&=-2n(n-1)h_{kl}{}^{(0)\<0\>}_{010}h^{ik}{}^{(0)\<0\>}_{000}h^{jl}{}^{(0)\<0\>}_{000}\\
&=-4\e K_{\zb}^{ij}{}_{00}^{(0)\<0\>}+\cO(\e^2),
\end{aligned}   
\end{equation}
Eq.~\eqref{eq:I0minusIanew_s4} becomes
\be\label{eq:s4L21K2no1}
    (I_0-I_a)^{(2)\<1\>} \supset
    \int\mathrm{d}^dy\,(\sqrt{h})^{(0)\<0\>}_{000}\bigg[
    -\frac{16\pi\e}{\kappa^2}K_{zij}{}_{00}^{(0)\<0\>}K_{\zb}^{ij}{}^{(0)\<0\>}_{00}+{\rm c.c.}\bigg]+\cO(\e^2).
\ee
Moving on to the second term in Eq.~\eqref{eq:spin4-L21}, we find that it only contributes through the non-Type I terms
\bal
   \(L_0\)^{(2)\<1\>}\supset\frac{1}{2}\(\sqrt{g}\)^{(0)\<0\>}_{000}
    \,4\(\phi_{zizj}{}^{(1)\<1\>}_{10-1}
    \phi^{zizj}{}^{(1)\<1\>}_{01-1}
    +\phi_{\zb i\zb j}{}^{(1)\<1\>}_{01-1}
    \phi^{\zb i\zb j}{}^{(1)\<1\>}_{10-1}\)
    (z\zb)^{n-2}
    ,
\eal
which gives
\bal
\hspace{-2mm}    (I_0\,{-}\,I_a)^{(2)\<1\>}{\supset}\! \int \mathrm{d}^dy\,(\sqrt{h})^{(0)\<0\>}_{000}\[\frac{2\pi}{\e} \phi_{zizj}{}^{(1)\<1\>}_{10-1} \phi^{zizj}{}^{(1)\<1\>}_{01-1}{+}{\rm c.c.}\]\!{+}\,\cO(\e^2).
\eal
Again using Eq.~\er{eq:phi11s4}, this expression becomes
\bal\label{eq:s4L21K2no2}
\hspace{-2mm}   (I_0-I_a)^{(2)\<1\>}&{\supset}\!\int \!\!\mathrm{d}^dy (\sqrt{h})^{(0)\<0\>}_{000}\[ \frac{2\pi}{\k^2 \e} R_{zizj}{}^{(0)\<0\>}_{10-1} R^{zizj}{}^{(0)\<0\>}_{01-1}{+}{\rm c.c.}\!\]\!{+}\,\cO(\e^2).\\
    &{=}\!\int\!\! \mathrm{d}^dy(\sqrt{h})^{(0)\<0\>}_{000}\[\frac{8\pi\e}{\k^2}K_{zij}{}^{(0)\<0\>}_{00} K_{\zb}^{ij}{}^{(0)\<0\>}_{00}\times 2 \]+ \cO(\e^2).
\eal
Again, at this order in $\a$ and $1/m^2$, we have no further contributions. Combining the results from Eqs.~\eqref{eq:s4L21K0}, \eqref{eq:s4L21K2no1}, and \eqref{eq:s4L21K2no2}, we have
\be 
\begin{aligned}
    &(I_0-I_a)^{(2)\<1\>}   \supset\\
    &\, 2\pi\e \int \mathrm{d}^dy\,(\sqrt{h})^{(0)\<0\>}_{000}\bigg[
    \frac{1}{\kappa}\left(\phi^{z\zb\zb z}{}_{000}^{(1)\<1\>}
    + \phi^{z\zb\zb z}{}_{11-1}^{(1)\<1\>}\right)
     -\frac{8}{\kappa^2}K_{zij}{}_{00}^{(0)\<0\>}K_{\zb}^{ij}{}_{00}^{(0)\<0\>} \, \bigg] \\& +\cO(\e^2).
\end{aligned}
\ee 
Adding this result to the area contributions given in Eqs.~\eqref{eq:s2L0einstein}, \eqref{eq:s2L1einstein}, and \eqref{eq:s2L2einstein}, we find
\be\label{eq:Agen_spin4}
\hspace{-4.5mm}    A_\gen = \frac{2\pi}{\kappa^2} \int \mathrm{d}^dy\sqrt{h} \left[1 + \frac{\alpha^2}{m^2}\left(\kappa\phi^{z\zb\zb z}{}^{(1)\<1\>} - 8K_{zij}{}^{(0)\<0\>}K_{\zb}^{ij}{}^{(0)\<0\>}\right)\right] + \cO\(\fr{\a^4}{m^4}\)\hspace{-1mm}
\ee
up to higher order terms in $1/m^2$.
Writing this covariantly in $z$, $\zb$ and using $\phi^{ab}{}_{ab}{}^{(0)}=0$, we find
\be
A_\gen = \frac{2\pi}{\kappa^2} \int {\rm d}^d  y \sqrt{h} \left[ 1 +  2 \alpha \kappa \phi^{ab}{}_{ab} -\frac{2\alpha^{2}}{m^{2}}K_{aij}{}K^{aij}{} \right]+\cO\(\fr{\a^4}{m^4}\),
\ee
which is precisely Eq.~\eqref{eq:Agenfinal4}.

\section{Holographic entanglement entropy under field redefinition}
\label{sec:LM_redef}

A particular representation of an EFT Lagrangian, in terms of higher-dimension interactions and their associated Wilson coefficients, is only well-defined---in the sense of being uniquely specified given IR physics---modulo field redefinition.
For example, Einstein gravity $\cL\sim R$ can be field redefined to $\cL\sim R+\lambda_1 R^2 + \lambda_2 R^{\mu\nu}R_{\mu\nu}$ (to leading order in the Wilson coefficients $\lambda_1,\lambda_2$), so the two theories are in fact equivalent.\footnote{Note that the $R^{\m\n\r\s}R_{\m\n\r\s}$ term is not related to Einstein gravity via a field redefintion. Hence, we do not consider it in this appendix.} 
A priori, there might be some tension: if two bulk theories are equivalent, the dual boundary theory should be the same, and therefore the boundary entropy should be independent of which bulk description we use; on the other hand, the two equivalent theories do have different Lagrangians leading to potentially different entropy functionals, and one might question whether they have the same extremal value.

In this appendix, we will see that the two representations of the theory do in fact give the same entanglement entropy, despite apparent dissimilarities. We will demonstrate this in two simple examples. The key ingredients in the resolution are the equations of motion and the extremality condition.

\subsection{From Einstein to \texorpdfstring{$R^2$}{R2}}
To demonstrate this equivalence with an example, consider the Euclidean action
\begin{equation}
\label{eq:action_Ein}
I_\diamond = \frac{1}{2} \int \mathrm{d}^{D} x \sqrt{g_{\diamond}}\left(-R_{\diamond}+2 \Lambda+g_{\diamond}^{\mu \nu} \nabla_{\mu} \phi \nabla_{\nu} \phi\right)
\end{equation}
where the subscript $\diamond$ denotes the quantities before redefinition and $D=d+2$ is the bulk spacetime dimension. We have included a scalar field $\p$ to make the example more nontrivial.
The entropy functional is simply the area
\begin{equation}
\label{eq:unredef_area}
A_{\gen,\diamond} =2 \pi \int \mathrm{d}^{d} y \sqrt{h_{\diamond}}.
\end{equation}
Einstein's equation is given by the vanishing of
\begin{align}\la{Einseq}
E_{\diamond}^{\mu \nu} &\equiv \frac{2}{\sqrt{g_{\diamond}}} \frac{\delta I}{\delta g_{\diamond \mu \nu}}=R_{\diamond}^{\mu \nu}-\frac{1}{2} R_{\diamond} g_{\diamond}^{\mu \nu}+\frac{1}{2}\left(2 \Lambda+\nabla_{\mu} \phi \nabla^{\mu} \phi\right) g_{\diamond}^{\mu \nu}-\nabla^{\mu} \phi \nabla^{\nu} \phi,
\end{align}
where it is understood that the index on $\nabla^\mu$ is raised with $g_\diamond$.

Now let us perform a small field redefinition, 
\begin{equation}
g_{\diamond \mu \nu}=(1+\lambda R) g_{\mu \nu}, \quad \delta g_{\mu \nu} \equiv g_{\diamond \mu \nu}-g_{\mu \nu}=\lambda R g_{\mu \nu}\label{eq:FD1}
\end{equation}
and work to linear order in $\lambda$. The action becomes 
\be 
\begin{aligned}
    I&=\frac{1}{2} \int \mathrm{d}^{D} x \sqrt{g}\left[-R+2 \Lambda+(\nabla \phi)^{2}\right]+\frac{1}{2} \int \mathrm{d}^{D} x \sqrt{g} E^{\mu \nu} \delta g_{\mu \nu}\\
    &=\frac{1}{2} \int \mathrm{d}^{D} x \sqrt{g} \left[-R+2 \Lambda+(\nabla \phi)^{2} +\l\(-\frac{d}{2} R^2+(d+2) \Lambda R+\frac{d}{2}(\nabla \phi)^{2}R\)\right],
\end{aligned}
\ee
up to boundary terms that are not important for our purposes. Using the method in Sec.~\re{sec:met}, we find the entropy functional $A_\gen$ for this new action:
\begin{equation}\la{Agennew}
A_\gen = 2 \pi  \int \mathrm{d}^{d} y \sqrt{h}\left[1+\lambda\left(d R-(d+2) \Lambda-\frac{d}{2}(\nabla \phi)^{2}\right)\right].
\end{equation}
On the other hand, if we simply apply the field redefinition~\eqref{eq:FD1} to the entropy functional~\eqref{eq:unredef_area}, we find
\begin{equation}
A_\gen' =2 \pi \int \mathrm{d}^{d} y \sqrt{h}\left(1+ \fr{d}{2} \l R\right),
\end{equation}
which appears to disagree with \er{Agennew}.

To resolve this discrepancy, we need to use the equations of motion~\er{Einseq}. Specifically, we only need the trace,
\begin{equation}
E_{\mu}^{\mu} = -\frac{d}{2} R + (d + 2) \Lambda + \frac{d}{2}(\nabla \phi)^{2}=0.
\end{equation}
We then have
\begin{equation}
A_\gen = 2 \pi  \int \mathrm{d}^{d} y \sqrt{h}\left(1 + \fr{d}{2} \l R\right),
\end{equation}
so the expressions agree: $A'_{\gen} = A_\gen$. 

In this example, we have seen that the equations of motion alone are enough to equate two apparently different expressions for the entropy functional. Next, we consider an example where this is not the case.

\subsection{From Einstein to \texorpdfstring{$R_{\mu\nu}R^{\mu\nu}$}{Ricci2}}
Consider again Einstein gravity with a cosmological constant \eqref{eq:action_Ein}, but this time for simplicity without the matter field $\p$. To generate a Ricci-squared term, we take our field redefinition to be
\begin{equation}
\label{eq:redef2}
    g_{\diamond \mu\nu}=g_{\mu\nu}+\lambda R_{\mu\nu}.
\end{equation}
Then the action becomes
\begin{align}
    I=\frac{1}{2} \int \mathrm{d}^{D} x \sqrt{g}\left[-R+2 \Lambda+ \lambda \left(\Lambda R-\frac{1}{2}R^2+R_{\mu\nu}R^{\mu\nu}\right)\right].
\end{align}
Applying Eq.~\er{eq:S_Dong13} to this new action, we find the entropy functional to linear order in $\l$:
\begin{equation}
\label{eq:S1_ricci}
    A_\gen = 2\pi \int \mathrm{d}^{d}y \sqrt{h}\left[1+\lambda\left(- \Lambda+R-R^a_a
    +\frac{1}{2}K^a K_a\right)\right].
\end{equation}
Applying the field redefinition \eqref{eq:redef2} directly to the entropy functional \eqref{eq:unredef_area}, on the other hand, yields
\begin{equation}\la{Agenprime}
    A_\gen' = 2\pi\int \mathrm{d}^{d}y \sqrt{h}\left(1+\fr{\lambda}{2} R^i_i \right).
\end{equation}
To check whether this agrees with \er{eq:S1_ricci}, we again use the equations of motion. Taking the trace of the metric equation of motion in all directions, in the $a$-type directions, and in the $i$-type directions gives, respectively,
\begin{align}
    R = \frac{2(d+2)\Lambda}{d} , \qqu
    R^a_a = \frac{4\Lambda}{d} , \qqu
    R^i_i = 2\Lambda.
\end{align}
Substituting these equations into Eqs.~\eqref{eq:S1_ricci} and \er{Agenprime}, we obtain
\be 
\begin{aligned}
    A_\gen &= 2\pi\int \mathrm{d}^{d}y \sqrt{h}\left[1+\lambda\left(\Lambda+\frac{1}{2}K^aK_a\right)\right],\\
    A_\gen' &= 2\pi\int \mathrm{d}^{d}y \sqrt{h}\left(1+\lambda  \Lambda \right).
\end{aligned}
\ee
The difference between $A_\gen$ and $A_\gen'$ is now given by a term proportional to $K^aK_a$. This is where we need to use the extremality condition at zeroth order in $\lambda$, $K_a=0$, so the extra term $K^aK_a$ vanishes (up to higher order terms in $\lambda$). Thus, we have shown that even though the entropy functional of the field-redefined theory generally does not match the field-redefined entropy functional, their extremal values (i.e., the entropy values) are indeed the same.

\end{appendix}

\bibliographystyle{JHEP}
\bibliography{bibliography}

\end{document}